\documentclass[11pt]{article}
\usepackage{amssymb}
\usepackage{amsmath}
 \newcommand{\N}{\mathbb{N}}
 \newcommand{\R}{\mathbb{R}}
 \newcommand{\C}{\mathbb{C}}

 \newcommand{\df}{\mbox{\,{\rm :}$=$\,}}
 \newcommand{\Bix}{\rule{0.6em}{0.6em}}
 \newcommand{\frk}[2]{\mbox{$\frac{#1}{#2}$}}
 \newcommand{\cB}{{\cal{B}}}
 \newcommand{\cC}{{\cal{C}}}
 \newcommand{\cD}{{\cal{D}}}
 \newcommand{\cH}{{\cal{H}}}
 \newcommand{\cK}{{\cal{K}}}
 \newcommand{\cL}{{\cal{L}}}
 \newcommand{\cO}{{\cal{O}}}
 \newcommand{\cS}{{\cal{S}}}
 \newcommand{\cW}{{\cal{W}}}
 \newcommand{\dA}{{\mathfrak{A}}}
 \newcommand{\vk}{{\vec{k}}}
 \newcommand{\vu}{{\vec{u}}}
 
 \newcommand{\vx}{{\vec{x}}}
 \newtheorem{prop}{Proposition}[section]
 \newtheorem{lem}[prop]{Lemma}

\title{On Infravacua and the Localisation of Sectors}
\author{Walter Kunhardt
     \\ Institut f\"ur Theoretische Physik der Universit\"at G\"ottingen
     \\ Bunsenstra\ss{}e 9, 37073 G\"ottingen, Germany
     \\ e-mail: {\tt kunhardt@theorie.physik.uni-goettingen.de} }
\date{June 5, 1998}

\begin{document}
\maketitle
\begin{abstract}
A certain class of superselection sectors of the free massless scalar
field 
in 3 space dimensions is considered. It is shown that these sectors,
which cannot be localised with respect to the vacuum, acquire a much
better localisation, namely in spacelike cones, when viewed in front
of suitable ``infravacuum'' backgrounds. These background states
coincide, essentially, with a class of states introduced by Kraus,
Polley and Reents as models for clouds of infrared radiation.
\end{abstract}

\section{Introduction}
In the analysis of superselection sectors, the localisability
properties of charges are crucial for defining notions as charge
composition and conjugation, statistics or a (global) gauge group. In
a classical work \cite{DHR}, Doplicher, Haag and Roberts carried
through such a programme for charges which fulfill what is now called
the DHR criterion, i.e., which are compactly localised. Now this
criterion is very restrictive, and Buchholz and Fredenhagen \cite{BF82}
established that sectors of theories in 3 space dimensions
without massless particles in general only comply with the so-called
BF criterion, i.e., they are localised in spacelike cones. Still,
these authors could extend the analysis of \cite{DHR} to charges with
such a weaker localisation behaviour.

The situation is more difficult for theories with massless
particles. Typically, these theories possess sectors whose
localisation is too poor for the DHR framework to be applicable.
Motivated by what
is expected to happen in QED, it has been proposed by Buchholz in
\cite{Bu82} to improve the localisation by viewing the charges in front
of some suitable background field instead of the vacuum. In QED,
such background fields should correspond to clouds of infrared
radiation. An appropriate mathematical description of such infrared
clouds has been introduced by Kraus, Polley and Reents in \cite{KPR}.

Here, we want to verify this mechanism in a simpler
model, namely in the theory of the free massless scalar field in 3
space dimensions \cite{BDMRS}. More precisely, we will consider a certain
class of (non-Lorentz invariant) sectors described by automorphisms of the
observable algebra and analyze their localisation properties in terms
of the BF criterion. In particular,
we will show that the sectors under consideration do satisfy this
criterion with respect to a KPR-like  background but do not
satisfy it with respect to the vacuum. (Calling the background fields
``KPR-like'' should indicate that they are very similar, yet not
identical, to those of \cite{KPR}.)

As to the consistency of such an approach, it should be kept in mind
that, in a theory whose charges are compactly localised, the
superselection
structure can be described without any difference with respect to the
vacuum as well as with respect to so-called infravacua, the latter
being generalisations of the KPR-like background states considered
here. As has been shown in \cite{WK}, when viewed in front of such an
infravacuum, the charges remain compactly localised and have the same
fusion structure and statistics as in front of the vacuum. Moreover,
positivity of the energy in a sector does not depend on the background
chosen, nor do the masses of massive particles possibly contained in
such a theory.

The above-mentioned class of sectors of the free massless field
has been studied recently by
Buchholz et al.\ \cite{BDMRS} with the purpose of modeling charges
of electromagnetic type. In the following, we will stick very closely
to the notations introduced there, but we should emphasize that our
point of view is slightly different from that adopted in  \cite{BDMRS}:
Buchholz et al.\  achieved a better localisation of the sectors by
restricting them to a (non-Lorentz invariant) subnet $\dA_0 \subset \dA$
of the observable net. The sectors then even became localised in the
DHR sense, which permitted them to carry through a DHR-like analysis, 
even though the net $\dA_0$ does not fulfill Haag duality. Here, in
contrast, the subnet $\dA_0$ will play no r\^{o}le, and the
localisation obtained will be in a weaker sense.

We end this Introduction by recalling the definition of the model
under consideration. The observable algebra of the free massless
scalar field is defined in its vacuum representation. More precisely,
let $\cK\df L^2(\R^3, d^3k)$ be the Hilbert space of momentum space
wave functions, $\omega(\vk)\df|\vk|$ the one-particle energy
and $U(t,\vx)= e^{i(\omega(\vk)t-\vk\vx)}$
the usual representation of the spacetime translations. The vacuum
Hilbert space of our model will be the bosonic Fock space $\cH$ over
$\cK$; the induced unitary representation of the spacetime
translations will still be denoted by $U(t,\vx)$ without any risk of
confusion. For any $v\in\cK$, $W(v)\in \cB(\cH)$ will denote the
corresponding Weyl operator. The normalisation is chosen such that the
Weyl relations read
$W(u)W(v)=e^{-\frac{i}{2}\mbox{\footnotesize ${\rm Im}$}\langle
u,v\rangle}W(u+v)$.
For any real linear subspace $\cL\subset\cK$, $\cW(\cL)$ denotes the
C$^*$-subalgebra of $\cB(\cH)$ generated by the operators $W(f)$,
$f\in\cL$. The net of observables now is given as
$$ \cO \longmapsto \dA(\cO) \df \cW(\cL(\cO))'', $$
where $\cO\longmapsto\cL(\cO)$ is the isotonous, local and covariant net
of
symplectic subspaces in $\cK$ (indexed by the set of open double cones
in Minkowski space) defined as follows: If $\cO\df (\{0\} \times O)''$
is the causal completion of an open ball $O\subset\R^3$ at time $t=0$, one
has
$$ \cL(\cO) \df \omega^{-\frac{1}{2}} \widehat{\cD_\R(O)}
{            }+i\omega^{+\frac{1}{2}} \widehat{\cD_\R(O)} , $$
where $\cD_\R(O)$ is the set of all real-valued
smooth functions with support in
$O$ and $\hat{\mbox{ }}$ denotes the Fourier transform. For other double
cones $\cO$, the space $\cL(\cO)$ is defined by
translation covariance and additivity. The
symplectic form $\sigma$ on $\cL\df\bigcup_\cO \cL(\cO)\subset\cK$ reads
$$ \sigma(f_1,f_2) \df -{\rm Im}\langle f_1,f_2 \rangle ,$$
and locality for the net $\cL(\cdot)$ just means
$\sigma(\cL(\cO_1),\cL(\cO_2))=0$ whenever  $\cO_1$ and $\cO_2$ are
spacelike to each other.
As usual, we also associate symplectic subspaces of $\cL$
(resp.\  C$^*$-subalgebras of $\cB(\cH)$) to {\em un}bounded regions in
$\R^{1+3}$
by additivity (resp.\ additivity and norm closure) and simply denote
by $\dA$ the quasilocal algebra $\dA(\R^{1+3})$.

The charges under consideration are given in terms of net automorphisms
$\gamma\in{\rm Aut}\dA$  which are labeled uniquely by  elements of the
(additive) abelian group
$$ \cL_\Gamma \df \omega^{-\frac{1}{2}} \widehat{\cD_\R(\R^3)}
{          }+i\omega^{-\frac{3}{2}} \widehat{\cD_\R(\R^3)} . $$
Any element $\gamma\in\cL_\Gamma$ gives rise to a linear form
$l_\gamma:\cL\longrightarrow\C$,
$$ l_\gamma(f) \df -{\rm Im} \int d^3k \; \overline{\gamma(\vk)} \: f(\vk)
$$
and hence to an automorphism, again denoted by $\gamma$, of $\dA$ by
$$ \gamma(W(f)) \df e^{il_\gamma(f)} W(f)  . $$
As explained in \cite{BDMRS}, $\gamma$ is indeed a well-defined
automorphism of the quasi-local algebra $\dA$ since, by Huygens'
principle, it turns out to be locally normal; as a consequence, it can
be extended by weak continuity from the local Weyl algebras
$\cW(\cL(\cO))$
to the local von Neumann algebras $\dA(\cO)$.

There will be no risk of confusion in viewing the real vector space
$\cL_\Gamma$ as an abelian subgroup of ${\rm Aut}\dA$. In particular, a
sum $\gamma_1+\gamma_2$ in $\cL_\Gamma$ corresponds to the composition
$\gamma_1\circ\gamma_2$ in ${\rm Aut}\dA$. Moreover, $\gamma_1$ and
$\gamma_2$
define
the same sector of $\dA$,  i.e.,  they are unitarily equivalent in
$\cB(\cH)$,
iff $\gamma_1-\gamma_2\in \cL_\Gamma\cap\cK$. In this case, the Weyl
operator
$W(\gamma_1-\gamma_2)$ is well defined  and implements the unitary
equivalence
$\gamma_1\cong\gamma_2$ on $\dA$.

Any $\gamma\in \cL_\Gamma$ can be written uniquely in the form
$\gamma= \omega^{-\frac{1}{2}} \hat{\sigma} + i\omega^{-\frac{3}{2}}
\hat{\rho}$
with functions $\sigma,\rho \in \cD_\R(\R^3)$. Since $\hat{\sigma}$ and
$\hat{\rho}$
are analytic, it is obvious that $\gamma$ is square integrable, i.e.,
$\gamma\in\cK$, iff $\hat{\rho}(0)=0$. As a consequence, the sectors
considered are labeled by a single real parameter
$$ q_\gamma \df \hat{\rho}(0) = \int d^3x\;\rho(\vx) $$
which is interpreted as the charge of the sector $[\gamma]$. In
particular, this shows that the sectors are transportable;
as a matter of fact, they even have positive energy \cite{BDMRS}.

\section{Bad localisation of the sectors in front of the vacuum}
It has been shown in \cite{BDMRS} that the automorphisms
$\gamma\in\cL_\Gamma$
do not satisfy the DHR localisation criterion. Here, we want to strengthen
this
result and show with closely related methods that they do not even
satisfy the BF criterion, that is, that they are not localisable in
spacelike cones. To this end, it is sufficient to prove the
following

\begin{prop}
Let $C\subset\R^3$ be an open convex cone having $0$ as its apex and
denote
with $\cC\df (\{0\}\times C)''$ its causal completion.
Then, for any $\gamma\in \cL_\Gamma$,
$$\gamma|_{\dA(\cC)}\cong {\rm id}|_{\dA(\cC)} \quad {\rm iff} \quad
q_\gamma
=0.$$
\end{prop}

The ``if'' part of this proposition is trivial, and before proving the
``only if'' part, we recall some facts about the dilation covariance
of the model. The dilation group $\R_{>0}$ acts unitarily on $\cK$ and
leaves the space $\cL$ invariant. More precisely, $f\in\cL(\cO)$ is
mapped onto $f_\lambda \in\cL(\lambda\cO)$, where
$f_\lambda(\vk)\df \lambda^\frac{3}{2} f(\lambda\vk)$. Writing
$f=\omega^{-\frac{1}{2}}\hat{h}+i\omega^{+\frac{1}{2}}\hat{g}$, it is
verified
by a
straightforward computation that this entails for the linear form
$l_\gamma$, $\gamma\in \cL_\Gamma$ :
$$ l_\gamma(f_\lambda) =
      \int
\frac{d^3k}{\omega^2}\,\overline{\hat{\rho}(\vk/\lambda)}\,\hat{h}(\vk)
     -\frac{1}{\lambda} \int
d^3k\,\overline{\hat{\sigma}(\vk/\lambda)}\,\hat{g}(\vk).  $$
In the limit $\lambda\to\infty$,
$\vk\longmapsto \overline{\hat{\rho}(\vk/\lambda)}\hat{h}(\vk)$ converges
to $\hat{\rho}(0)\hat{h}$ in the space of test functions, and since
$\frac{2\pi^2}{r}$ is the Fourier transform of $\frac{1}{\omega^2}$ in the
sense of distributions, one obtains
$$ \lim_{\lambda\to\infty} l_\gamma(f_\lambda) = q_\gamma \: \kappa_f
\quad
{\rm with} \quad
     \kappa_f\df 2\pi^2 \: \int \frac{d^3 x}{|\vx|}\, h(\vx)  .$$
This allows us to prove the following
\begin{lem}
Let $f\in\cL(\cO')$, where $\cO\subset\R^{1+3}$ is a neighbourhood of $0$.
For
any $\gamma\in\cL_\Gamma$, one then has
$$ \mbox{\rm
w-}\!\lim_{\!\!\!\!\!\!\!\lambda\to\infty}\gamma(W(f_\lambda))
     = e^{iq_\gamma \, \kappa_f}  \: e^{-\frac{1}{4}\|f\|^2}\:{\bf 1}.  $$
\end{lem}
Proof: Since the dilations act geometrically, it follows by locality
from the special form of the localisation region of $f$ that
$\lim_{\lambda\to\infty}\sigma(f_\lambda,f')=0$ for any $f'\in\cL$. Hence,
$(W(f_\lambda))_{\lambda>0}$ is a central sequence of unitaries in
$\cW(\cL)$
whose set of weak limits is, by the irreducibility of the vacuum  
representation, a (nonempty) subset of $\C\:{\bf 1}$.  On the other
hand, unitarity of the dilations permits us to evaluate this limit in the
vacuum state: $\omega_0(W(f_\lambda)) \smash{\stackrel{\lambda\to\infty}
{\smash{-}\!\!\smash{-}\!\!\!\longrightarrow}} e^{-\frac{1}{4}\|f\|^2}$.
But this means that $W(f_\lambda)$ has $e^{-\frac{1}{4}\|f\|^2}\,{\bf 1}$
as
its {\em unique} weak limit for $\lambda\to\infty$, establishing thus
the assertion for $\gamma=0$. For arbitrary $\gamma\in\cL_\Gamma$, it now
follows
easily in view of the discussion in the preceding paragraph.
\Bix

Physically, the sequence $(W(f_\lambda))_{\lambda\to\infty}$ is
interpreted as
a
measurement of the asymptotic behaviour (in the spatial directions
determined by the smearing function $h$) of the ``Coulomb potential''
of the ``charge density'' $\rho$. In QED, one expects that operators
measuring the asymptotic electric flux distribution play a similar
r\^{o}le, cf.\ \cite{Bu82}. In the present  case, the leading $1/r$
behaviour of the Coulomb potential is isotropic in all sectors
$[\gamma]$. This fact, reflected by the factorizing of $\lim
l_\gamma(f_\lambda)$
as seen above, is relevant in the\\
Proof of Prop.~2.1:
Let $\gamma\in\cL_\Gamma$ with $q_\gamma\neq 0$. Choose a nonvanishing,
nonnegative test function $h\in\cD_\R(C)$. Letting
$f\df \omega^{-\frac{1}{2}}\hat{h}$, this implies $\kappa_f\neq 0$ and
$f\in\cL(\cC\cap\cO')$ for some neighbourhood $\cO\subset\R^{1+3}$ of
$0$. Since $e^{iq_\gamma \, \kappa_f} \neq 1$ can always be achieved by a
mere
rescaling of $h$, Lemma~2.2 shows that the weak limits (as
$\lambda\to\infty$)
of $W(f_\lambda)$ and $\gamma(W(f_\lambda))$ are different scalar
multiples of
the
unit operator. But since $W(f_\lambda)\in \dA(\cC)$ for all $\lambda > 0$,
this
implies $\gamma|_{\dA(\cC)} \not\cong {\rm id}|_{\dA(\cC)} $.
\Bix

\section{Infravacuum background states}
In this section we introduce a class of background states in front of
which the automorphisms $\gamma$ will be shown (in Section~4) to
have better localisation properties. Apart from two modifications
necessitated by the present model, these background states are of the same
type
as those introduced by Kraus, Polley and Reents \cite{KPR} as a model
for infrared clouds in QED or, more generally, in any theory
containing massless particles.

\subsection{Preliminaries on quasifree states}
First, we recall that a {\em quasifree state} on $\dA$ is a locally
normal state $\omega_T$ which is, on the Weyl operators $W(f)\in\dA$,
$f\in\cL$
of the form
$$ \omega_T(W(f)) = e^{-\frac{1}{4}\|Tf\|^2}  .   $$
Here, $T:D_T\longrightarrow\cK$ is a real linear, symplectic (i.e.,
fulfilling
${\rm Im} \langle Tv, Tw \rangle = {\rm Im} \langle v, w \rangle$,$v,w\in
D_T$)
operator defined on a dense, real linear subspace $D_T$ which
contains $\cL$. In the case at hand, we will have in addition
$\overline{T\cL} = \cK$, which entails that $\omega_T$ is a pure
state. Its GNS representation $\pi_T$ acts irreducibly on the vacuum
Hilbert space $\cH$ as $\pi_T(W(f))= W(Tf)$, $f\in\cL$.

Next, we describe the {\em real linear} operator $T$ in terms of a
pair of {\em complex linear} operators $T_1, T_2$ defined on complex
linear subspaces $D_{T_j}$, $j=1,2$.

\begin{lem}
Let $\Gamma:\cK\longrightarrow\cK$ be an antiunitary involution. Then, the
formulae
\begin{align*}
     T   & \df  T_2 \; \frk{1+\Gamma}{2} + T_1 \; \frk{1-\Gamma}{2}   \\
    D_T  & \df  \{ v\in \cK \mid \frk{1+\Gamma}{2}v\in D_{T_2},
                                \frk{1-\Gamma}{2}v\in D_{T_1}  \}
\end{align*}
establish a bijection between
\begin{itemize}
\item densely defined, $\Gamma$-invariant\footnote{
              Here, $T:D_T\longrightarrow \cK$ being $\Gamma$-invariant
means
              $\Gamma D_T= D_T$ and $[\Gamma,T]=0$ on $D_T$.}
 $\R$-linear operators $T:D_T\longrightarrow\cK$
   and
\item densely defined, $\Gamma$-invariant $\C$-linear operators
   $T_j:D_{T_j}\longrightarrow\cK$, \\ $j=1,2$.

\end{itemize}
Moreover, $T$ is symplectic iff
$\langle T_1 u_1, T_2 u_2 \rangle = \langle u_1, u_2 \rangle$ for all
$u_j\in D_{T_j}$.
\end{lem}

Since all assertions can be checked by simple calculations, we omit
the formal proof of this Lemma and merely point out that the converse
formulae expressing $T_1$ and $T_2$ in terms of $T$ read
\begin{eqnarray*}
 & D_{T_2} =  \{ v\in \cK \mid \frk{1+\Gamma}{2}\C v \subset D_T \},
       \quad   & T_2 = T \; \frk{1+\Gamma}{2}  -i T \; \frk{1+\Gamma}{2}\,
i
\\
 & D_{T_1} =  \{ v\in \cK \mid \frk{1-\Gamma}{2}\C v \subset D_T \},
       \quad   & T_1 = T \; \frk{1-\Gamma}{2}  +i T \; \frk{1-\Gamma}{2}\,
i .
\end{eqnarray*}
{\bf Remark:} The involution $\Gamma$ induces the notion of
real and imaginary parts of vectors $v\in\cK$: ${\rm Re} v =
 \frac{1+\Gamma}{2}v $ , ${\rm Im} v = \frac{1-\Gamma}{2i}v$. Then, $T_2$
acts
on
 the real and $T_1$ on the imaginary parts:
$$ {\rm Re} \,Tv = T_2 \, {\rm Re} v , \quad {\rm Im} \, Tv = T_1 \,{\rm
Im} v,
\quad v\in D_T. $$

{}From now on, we will fix $\Gamma$ to be pointwise complex conjugation in
position space. In terms of momentum space wave functions $v\in\cK$,
this means
$$ (\Gamma v)(\vk) \:\df \: \overline{v(-\vk)}. $$
[For the sake of completeness, we point out that Kraus et al.\ used
pointwise conjugation in {\em momentum space} for defining their
background states in \cite{KPR}. In their case as well as in ours,
the choice of the involution $\Gamma$ is dictated by the set of sectors
under consideration.]

\subsection{Quasifree states with positive energy}
Before describing in detail the operators $T_1,T_2$, we introduce some
notation: for any $\epsilon>0$, let $P_\epsilon:\cK\longrightarrow\cK$ be
the
projector onto
the subspace $P_\epsilon \cK = \{v\in\cK | v(\vk)=0 \mbox{ if } |\vk| <
\epsilon \} $
and denote by
$$ D_0 \df \bigcup_{\epsilon>0} P_\epsilon \cK $$
the dense subspace of functions vanishing in some neighbourhood of
$\vk=0$. Note that $[P_\epsilon,\Gamma]=0$ and $\Gamma D_0 = D_0$. The
subspace
$D_0$
will serve as a provisional domain for $T_1$ and $T_2$.

Now we follow \cite{KPR} and choose
\begin{itemize}
\item a sequence $(\epsilon_i)_{i\in\N}$ in $\R_{>0}$ satisfying
  $\epsilon_{i+1} < \epsilon_i$ and
  $\epsilon_i\stackrel{i\to\infty}{-\!\!-\!\!\!\longrightarrow}0$.   \\
  This sequence induces a decomposition of momentum space into
  concentric spherical shells. The projections onto the associated
  spectral subspaces of $\cK$ will be denoted by
  $P_i\df P_{\epsilon_{i+1}}-P_{\epsilon_i}$. For notational convenience,
we
also
  put $P_0\df P_{\epsilon_1}$.
\item a sequence $(Q_i)_{i\in\N}$ of orthogonal projections in $\cK$ with
  finite rank ${\rm rk}Q_i$  satisfying $Q_i\Gamma = \Gamma Q_i$, $Q_i P_i
=
Q_i$.
\item a sequence $(b_i)_{i\in\N}$ in $]0,1[$ satisfying
  $ b_i\stackrel{i\to\infty}{-\!\!-\!\!\!\longrightarrow}0$ and
  $\sum_i \frac{\epsilon_i}{b_i^2}\,{\rm rk}Q_i <\infty$.   \\
  If, e.g., the $\epsilon_i$ decrease exponentially and ${\rm rk}Q_i$ is
  polynomially bounded, this can be satisfied by $b_i\propto
  i^{-\alpha}$, $\alpha >0$.
\end{itemize}
With these data, define $\C$-linear operators $T_1,T_2$ on the
subspace $D_0$ by
$$ T_1 \df 1+ \mbox{\rm s-}\!\lim_{\!\!\!\!\!\!n\to\infty}\sum_{i=1}^n
(b_i-1)Q_i,
\quad
   T_2 \df 1+ \mbox{\rm s-}\!\lim_{\!\!\!\!\!\!n\to\infty}\sum_{i=1}^n
                                    (\frac{1}{b_i} -1)Q_i.      $$
Since, on every $v\in D_0$, the number of terms which contribute on the
right-hand side is finite, these operators are well defined and
map $D_0$ into itself. Moreover, the relations
$$ T_1 P_i = ((1-Q_i) + b_i Q_i) P_i, \quad
   T_2 P_i = ((1-Q_i) +\frk{1}{b_i}  Q_i) P_i  $$
show that the subspace $P_i\cK$ decomposes into a subspace
$(1-Q_i)P_i\cK$ where both $T_1$ and $T_2$ act trivially and an
orthogonal subspace $Q_i P_i \cK = Q_i \cK$ where they act as
multiplications with the scalars $b_i$ and $\frac{1}{b_i}$,
respectively. As a consequence, $T_1$ and $T_2$ are inverses of each
other. Because of $\lim_{i\to\infty}b_i=0$, $T_1$ is bounded ($\|T_1\|
=1$), whereas $T_2$ is not. Also, it is clear that $T_1$ and $T_2$ are
$\Gamma$-invariant and symmetric. In particular, it follows that 
$\langle T_1 u_1,T_2 u_2 \rangle = \langle u_1, T_1 T_2 u_2 \rangle
=\langle u_1, u_2 \rangle $ for any $u_1,u_2\in D_0$. We are thus in
the situation of Lemma~3.1 and obtain an unbounded symplectic operator
$$ T:D_0\longrightarrow\cK,\quad T = T_2 \; \frk{1+\Gamma}{2} + T_1 \;
\frk{1-\Gamma}{2}. $$

In the next step, $T$ has to be extended to a larger domain
$D_T\supset \cL$. To this end, we analyze its singular behaviour
for $|\vk|\to 0$  by comparing it with powers of (a regularized
version $\omega_r$ of ) the one-particle Hamiltonian $\omega$. Setting
$$  \omega_r \df \omega\,(1-P_0)+ \epsilon_1\, P_0
      =  \left\{ \begin{array}{ll}
                  \omega   & \mbox{on  $\;(1-P_0)\cK$},   \\
                  \epsilon_1{\bf 1} & \mbox{on  $\;P_0\cK$}   \end{array}
\right.  $$
and noting that $\omega_r^{1/2} D_0 \subset D_0$, we obtain:
\begin{lem}
  $T_2\omega_r^{1/2}$ is bounded.
\end{lem}
Proof: Making use of $\|\omega_r P_i \|=\epsilon_i$ for $i\in\N$,
one obtains for $v\in D_0$
\begin{align*}
 \Big\|(T_2 -1)& \omega_r^\frac{1}{2} v \Big\|^2
   = \Big\| \sum_{i} (\frk{1}{b_i} -1) Q_i \omega_r^\frac{1}{2} v \Big\|^2
    = \sum_{i} (\frk{1}{b_i} -1)^2 \,
      \Big\langle \omega_r^\frac{1}{2} v, Q_i\:\omega_r^\frac{1}{2} v
\Big\rangle \\
   &\leq \sum_{i} (\frk{1}{b_i} -1)^2 \,{\rm rk}Q_i\,
            \Big\langle \omega_r^\frac{1}{2} v, P_i\:\omega_r^\frac{1}{2}
v
\Big\rangle
    \leq \sum_{i} (\frk{1}{b_i} -1)^2 \,{\rm rk}Q_i\, \epsilon_i\,
\|v\|^2.
\end{align*}
{}From the conditions imposed on the $b_i$, it follows that
$\sum_{i} (\frac{1}{b_i} -1)^2 \,{\rm rk}Q_i\,\epsilon_i $ is finite. Thus
$(T_2 -1)\omega_r^{1/2}$ is bounded, hence also $T_2 \omega_r^{1/2}$.
\Bix

We now can extend $T_1$ by continuity to all of $\cK =: D_{T_1}$ and
$T_2$ by the formula
$$ T_2 v \df T_2 \omega_r^\frac{1}{2} \, \omega_r^{-\frac{1}{2}} v, \quad
       v\in \omega_r^\frac{1}{2}\cK                               $$
to the dense subspace $\omega_r^\frac{1}{2}\cK  =: D_{T_2}$.
(Strictly speaking, the
symbol $ T_2 \omega_r^\frac{1}{2}$ on the right-hand side stands for the
continuous
extension to $\cK$ of the operator considered in the previous Lemma.)
Note that $T_1$ and $T_2$ still are $\Gamma$-invariant. We collect the
relevant properties in the following Lemma:

\begin{lem} {\quad}
  \begin{enumerate}
  \item $D_T\df \{v\in\cK | \frac{1+\Gamma}{2}v\in \omega_r^\frac{1}{2}\cK
\}$
    is a real linear dense subspace of $\cK$.
  \item $T = T_2 \: \frac{1+\Gamma}{2} + T_1 \: \frac{1-\Gamma}{2}$ is
    well defined on $D_T$.
  \item $T:D_T\longrightarrow\cK$ is a symplectic operator.
  \item $\cL\subset D_T$ and $T\cL$ is dense in $\cK$.
  \end{enumerate}
\end{lem}

Proof: Part 1 is obvious, since $D_0\subset D_T$; part 2 has been shown
in the previous paragraph. For 3, we have to show that
$\langle T_1 u_1,T_2 u_2 \rangle =\langle u_1, u_2 \rangle $
remains true for all $u_1\in D_{T_1}$ and $u_2\in D_{T_2}$. First,
assume $u_1\in D_0$. Since $D_0$ is dense in $\cK$ and invariant under
$\omega_r^{1/ 2}$, there exists a sequence $u_2^{(n)}\in D_0$, $n\in\N$
such
that
$ \omega_r{}\!\!^{-\frac{1}{2}} u_2 = \lim \omega_r{}\!\!^{-\frac{1}{2}}
u_2^{(n)}$, implying $u_2 = \lim u_2^{(n)}$. Using the boundedness
of $T_2 \omega_r^{1/2}$, we can compute
\begin{align*}
     \langle T_1 u_1,T_2 u_2 \rangle
  &=\langle T_1 u_1,T_2 \omega_r^\frac{1}{2} \, \omega_r^{-\frac{1}{2}}
u_2
\rangle
   = \langle T_1 u_1,T_2 \omega_r^\frac{1}{2} \, \lim_{n\to\infty}
                     \omega_r^{-\frac{1}{2}} u_2^{(n)}\rangle  \\
  &= \lim_{n\to\infty}  \langle T_1 u_1,T_2
                     \omega_r^\frac{1}{2} \, \omega_r^{-\frac{1}{2}}
u_2^{(n)}\rangle
   = \lim_{n\to\infty}  \langle  u_1, u_2^{(n)}\rangle
   = \langle  u_1, u_2 \rangle.
\end{align*}
Since $T_1$ is bounded, the restriction on $u_1$ can now be dropped by
continuity, thus yielding the assertion. Finally, $\cL\subset D_T$
is obvious, and the remaining part of 4 is equivalent,
in terms of $T_1$ and $T_2$, to
\begin{eqnarray*}
  \frk{1+\Gamma}{2} T\cL = T_2 \frk{1+\Gamma}{2}\cL
                      = T_2 \omega^{-\frac{1}{2}} \widehat{\cD_\R}
  & \mbox{\rm is dense in} &  \frk{1+\Gamma}{2}\cK           \\
  \frk{1-\Gamma}{2i} T\cL = T_1 \frk{1-\Gamma}{2i}\cL
                       = T_1 \omega^\frac{1}{2} \widehat{\cD_\R}
  & \mbox{\rm is dense in} &  \frk{1-\Gamma}{2i}\cK   .
\end{eqnarray*}
By $\C$-linearity, this in turn is equivalent to
$T_2 \omega^{-\frac{1}{2}} \widehat{\cD_\C}
=T_2 \omega_r^\frac{1}{2} \omega_r^{-\frac{1}{2}} \omega^{-\frac{1}{2}}
\widehat{\cD_\C}  $
and $T_1 \omega^\frac{1}{2} \widehat{\cD_\C}$ both being dense in $\cK$.
But this is implied by the fact that, on the one hand,
both operators $T_2 \omega_r^{1/2}$
and $T_1$ are bounded and have dense images (since they are
invertible on the dense, invariant subspace $D_0$) and that, on the
other hand,  the subspaces
$ \omega_r{}\!\!^{-\frac{1}{2}} \omega^{-\frac{1}{2}} \widehat{\cD_\C} $
and
$\omega^\frac{1}{2} \widehat{\cD_\C}$ are dense in $\cK$ (by the spectral
calculus
of $\omega$).
\Bix

With the above preparations, we can define a state
$\omega_T:\dA\longrightarrow\C$ and
analyze its main properties.
\begin{prop}
  The quasifree state $\omega_T$, defined on $\cW(\cL)$ by
$$ \omega_T(W(f)) = e^{-\frac{1}{4}\|Tf\|^2} , \quad  f\in\cL  $$
extends to a unique locally normal state $\omega_T$ over the quasilocal
algebra $\dA$. This state is pure and has positive energy.
\end{prop}

Proof: The difficult part of this proof is to obtain local normality
of $\omega_T$ on the net $\cO\longmapsto\cW(\cL(\cO))$ of Weyl algebras.
To
this end, recall that $T$ is  (on $\cL$) the strong limit of
symplectic operators $T_n$ such that $T_n-1$ have finite rank. As a
consequence, the associated quasifree states $\omega_{T_n}$ are vector
states in the vacuum representation and converge weakly to $\omega_T$ on
$\cW(\cL)$. Now since the Fredenhagen-Hertel compactness condition
C$_{\sharp}$ \cite{FrHe,BuPo86} is known to be fulfilled in the
present model, we can conclude that $\omega_T$ is locally normal if
the sequence $(\omega_{T_n})_{n\in\N}$ is bounded with respect to some
exponential energy norm $\|\cdot\|_\beta$, $\beta >0$ defined by
$\|\omega\|_\beta^2
\df \omega(e^{2\beta H})$. But this follows from
$\sum_i \frac{\epsilon_i}{b_i^2}\,{\rm rk}Q_i <\infty$, as F.\@ Hars
has shown in \cite{Hars}, adapting ideas from \cite{KPR}. (Although
our involution $\Gamma$ differs from that of \cite{KPR,Hars}, the
arguments leading to this conclusion are still valid.)
Hence, $\omega_T$ is locally normal on
$\cW(\cL)$ and thus extends uniquely to a locally normal state on
$\dA$. Since it is a weak limit of states in the vacuum
representation with positive energy, the arguments of
Buchholz and Doplicher \cite{BuDo84} can be applied to show that
$\omega_T$
has positive energy, too. Finally, the relation $\overline{T\cL}=\cK$,
established in Lemma~3.3, implies that $\omega_T$ is pure, as has been
noted at the very beginning of this section.
\Bix

{\bf Remark:}
The inequality  $\sum_i \frac{\epsilon_i}{b_i^2}\,{\rm rk}Q_i <\infty$,
which played a crucial r\^{o}le in the previous proof, has a direct
physical interpretation. Indeed, performing the limit
$\omega_{T_n}\to\omega_T$
corresponds to the excitation of more and more low-energy ``photon''
modes in comparison to the vacuum, namely those singled out by the
projections $Q_i$, $i=1,\dots,n$ which appear in $T$.
Since $\frac{1}{b_i}$ measures the amplitude of
these modes, each of them carries an energy of about
$\frk{\epsilon_i}{b_i^2}$. Hence the modes in the energy interval
$[\epsilon_{i+1},\epsilon_i]$
contribute with (at most) $\frac{\epsilon_i}{b_i^2}\,{\rm rk}Q_i$ to the
mean
energy
of the state $\omega_T$,  and the above inequality thus means that
$\omega_T$
describes an infrared cloud with finite total energy.
We conjecture that these arguments can be sharpened in order to prove
that the transition energy \cite{Wa87} between the sectors $\pi_0$ and
$\pi_T$
vanishes. In terms of \cite{WK}, the properties of $\pi_T$ could then
be summarised by saying that it is an ``infravacuum representation'', and
we will indeed use this terminology in the sequel.

\subsection{KPR-like quasifree states}
We reach our goal of improving the localisation of the automorphisms
$\gamma$ by considering a special class of infravacuum representations.
The main idea, due to \cite{KPR},  is to control the angular momentum
carried
by the low-energy modes. It may be formalised as follows.

{\bf Definition}: A state $\omega_T$ over $\dA$ based on the sequences
$\epsilon_i$,$Q_i$,$b_i$ as described above is called a KPR-like state
(and $\pi_T$ (resp.\ $T$) a KPR-like representation (resp.\  symplectic
operator)) if the following additional conditions are fulfilled:
\begin{enumerate}
\item $ (\ln \frac{\epsilon_i}{\epsilon_{i+1}})_{i\in\N}$ is polynomially
bounded, and
  $\sum_i b_i^2\,\ln \frac{\epsilon_i}{\epsilon_{i+1}} < \infty$.
\item With respect to the tensor product structure of the subspace
$P_i\cK$,
  $P_i\cK \cong L^2([\epsilon_{i+1},\epsilon_i], \omega^2d\omega) \otimes
L^2(S^2)$,
  the projections $Q_i$ read
  $$ Q_i = \frac{|\xi_i\rangle \langle \xi_i|}{\langle \xi_i|\xi_i\rangle
}
           \otimes \tilde{Q_i} \quad \mbox{with}\quad
     \tilde{Q_i}\df \sum_{0<l\leq i} \sum_{m=-l}^{l}
                  | Y_{lm}\rangle \langle Y_{lm} |   ;    $$
here the vector $\xi_i\in L^2([\epsilon_{i+1},\epsilon_i],
\omega^2d\omega)$ is
given by $\xi_i(\omega)
=\omega^{-\frac{3}{2}}$  and $Y_{lm}\in L^2(S^2)$ are the spherical
harmonics.
\end{enumerate}

This definition has been formulated so as to imply the regularity
property of the bounded operator $T_1$ formulated in the next
Lemma. It is only through this result that the two additional
properties of KPR-like infravacua enter the analysis of Section~4. It
is apparent from the ensuing proof that the above definition may be
generalised in several respects. However, we refrain from discussing these
possibilities here.

In contrast, we draw the reader's attention to the following crucial
difference between our KPR-like states and the ``true'' KPR states as
defined in \cite{KPR}: In our case, the projection $\tilde{Q_i}$ contains
no
summand $|Y_{00}\rangle \langle Y_{00}|$. In physical terms, this means
that the infrared cloud does not contain any spherically symmetric
low-energy modes. Such a restriction is necessary, since it is
precisely by such modes or, equivalently, by the isotropic long-range
behaviour of the ``Coulomb potential'', that the sectors $[\gamma]$ differ
from
each other. Too strong an $l=0$ contribution to the infrared cloud
would therefore render the sectors indistinguishable in front of that
background. (Indeed, if one had $0 \leq l \leq i$ in the definition of
$\tilde{Q_i}$, one would obtain, instead of Lemma~3.6 below, that
$\pi_T\circ\gamma \cong \pi_T$ for all $\gamma\in\cL_\Gamma$.) This
seemingly
artificial restriction on the background states mimics the situation
in QED, where the Coulomb field $\vec{\cal{E}}(\vk)\sim i\vk/{\omega^2}$
cannot be compensated by transverse photons.

\begin{lem}
  Let the sequences $\epsilon_i$,$Q_i$,$b_i$ be such that $\omega_T$ is a
  KPR-like state. Let $u\in\cK$ have, in a neighbourhood of $\vk=0$,
  the form $ u(\vk) = \eta({\vk}/{|\vk|}) $
  with some  $\eta\in C^\infty(S^2)\subset L^2(S^2)$. Then the sequence
  $(T_1\omega^{-\frac{3}{2}}P_{\epsilon_n}u)_{n\in\N}$ converges iff
$\eta\perp
Y_{00}$.
\end{lem}
Proof: Without any restriction, one may assume $u= c\otimes \eta $ with
$c(\omega)=1$ if $\omega<\epsilon_1$. Let $\eta\perp Y_{00}$. For $0<m<n$,
one
computes
\begin{align*}
  T_1\omega^{-\frac{3}{2}}P_{\epsilon_n}u \;-\;&
T_1\omega^{-\frac{3}{2}}P_{\epsilon_m}u
  \: =\: T_1\omega^{-\frac{3}{2}} \sum_{i=m}^{n-1} P_i\, (c \otimes \eta)
  \: =\: \sum_{i=m}^{n-1} T_1 P_i\,(\xi_i \otimes \eta)
\\=& \sum_{i=m}^{n-1} \,((1-Q_i) + b_i Q_i)(\xi_i \otimes \eta)
   = \sum_{i=m}^{n-1} \xi_i \otimes ((1-\tilde{Q_i})\eta + b_i \tilde{Q_i}
\eta).
\end{align*}
Now $\eta\in C^\infty(S^2)$ entails that
$\| (1-\tilde{Q_i})\eta \|^2 = \sum_{l>i} \sum_m |\langle
Y_{lm},\eta\rangle|^2$, $i\in\N$ is a sequence of rapid decrease (since
$\eta\in D(\vec{L}^{2\,N})$ for any $N$, $\vec{L}$ denoting the
angular momentum operator). Thus, using
$\| \xi_i\|^2 = \int_{\epsilon_{i+1}}^{\epsilon_i}
\omega^2\,d\omega\,\frac{1}{\omega^3}
  =\ln \frac{\epsilon_i}{\epsilon_{i+1}}$ and
$ \| b_i \tilde{Q_i} \eta \|^2\leq b_i^2\|\eta\|^2$, one obtains
\begin{eqnarray*}
  \Big\| T_1\omega^{-\frac{3}{2}}P_{\epsilon_n}u -
T_1\omega^{-\frac{3}{2}}P_{\epsilon_m}u\Big\|^2
  &\leq& \sum_{i=m}^{n-1} \|\xi_i\|^2 \, \Big( \| (1-\tilde{Q_i})\eta \|^2
                                          +b_i^2
\|\tilde{Q_i}\eta\|^2\Big)
\\
  &\leq& \sum_{i=m}^{n-1}\ln \frac{\epsilon_i}{\epsilon_{i+1}} \,
             \Big( \frac{c_N}{i^N} + b_i^2 \Big).
\end{eqnarray*}
With suitably chosen $N$, the right-hand side vanishes as
$m,n\to\infty$ due to the conditions imposed on $\epsilon_i$ and
$b_i$. Hence $(T_1\omega^{-\frac{3}{2}}P_{\epsilon_n}u)_{n\in\N}$ is a
Cauchy
sequence. Conversely, assume $\langle Y_{00},\eta \rangle \not= 0 $. With
$\eta = \langle Y_{00},\eta \rangle Y_{00} +\eta_1$,
$(T_1\omega^{-\frac{3}{2}}P_{\epsilon_n}(c\otimes \eta_1))_{n\in\N}$ is
convergent, hence $(T_1\omega^{-\frac{3}{2}}P_{\epsilon_n}u)_{n\in\N}$ is
divergent because
$(T_1\omega^{-\frac{3}{2}}P_{\epsilon_n}(c\otimes Y_{00}))_{n\in\N}
= (\omega^{-\frac{3}{2}}P_{\epsilon_n}(c\otimes Y_{00}))_{n\in\N}$ is.
\Bix

We end this section with a result which shows that the KPR-like
infravacua do not affect the superselection structure of the present
model. As the previous Lemma, it makes essential use of the fact that
$Tf=f$ for all rotation invariant elements $f\in D_T$.

\begin{lem} Let $\pi_T$ be a KPR-like infravacuum
  representation. Then, for any $\gamma_1,\gamma_2\in\cL_\Gamma$, one has
$$ \pi_0\circ\gamma_1 \cong \pi_0\circ\gamma_2
   \quad\mbox{iff} \quad  \pi_T\circ\gamma_1 \cong \pi_T\circ\gamma_2 .
$$
\end{lem}

Proof: Let $\pi_0\circ\gamma_1 \cong \pi_0\circ\gamma_2$. Then
$\gamma\df\gamma_1-\gamma_2\in\cL_\Gamma$ has charge $q_\gamma=0$, as
noted in
the
Introduction, which does not only yield $\gamma\in\cK$, but even
$\gamma\in
D_T$. Hence, the unitary $W(T\gamma)$ is well defined and intertwines the
representations $\pi_T\circ\gamma_1$ and $ \pi_T\circ\gamma_2$.
Conversely,
assume $\pi_0\circ\gamma_1 \not\cong \pi_0\circ\gamma_2$, i.e.,
$q_{\gamma_1}\not=q_{\gamma_2}$. For any {\em rotation invariant} test
function $h\in\cD_\R(\R^3\setminus\{0\})$, one has
$\omega^{-\frac{1}{2}}\hat{h} \mbox{\,$=${\rm :}\,}f \in \cL(\cO')$ for
some open neighbourhood $\cO\subset\R^{1+3}$ of $0$. Since
$Tf_\lambda=f_\lambda$,
Lemma~2.2 implies
$$ \pi_T\circ\gamma_j(W(f_\lambda)) = \gamma_j(W(f_\lambda))
   \stackrel{\lambda\to\infty}{-\!\!-\!\!\!\longrightarrow}
   e^{iq_{\gamma_j}\kappa_f}\: e^{-\frac{1}{4}\|f\|^2}\:{\bf 1}.       $$
As it is always possible to obtain
$e^{iq_{\gamma_1}\kappa_f}\neq e^{iq_{\gamma_2}\kappa_f}$ by a rescaling
of
$h$, it
follows that $\pi_T\circ\gamma_1\not\cong\pi_T\circ\gamma_2$.
\Bix

\section{Better localisation of the sectors in front of KPR-like
 infravacua}
The main aim of this section is to prove the following result which
establishes some (non-Lorentz invariant) version of BF
localisation. In the sequel, we will denote by  $\cC=(\{t\}\times C)''$
an ``upright'' spacelike cone whose basis is the open convex cone
$C\subset\R^3$ at time $t$. Note that the set of upright spacelike cones
is translation invariant and that an arbitrary spacelike cone can be
obtained from an upright one by a Lorentz transformation.

\begin{prop}
  Let $\pi_T$ be a KPR-like infravacuum representation, and let
  $\gamma\in\cL_\Gamma$. Then one has for any upright spacelike cone
$\cC$:
$$ \pi_T\circ\gamma|_{\dA(\cC')} \cong \pi_T |_{\dA(\cC')}. $$
\end{prop}

To prove this assertion, we will first deal with a special case in
which the relevant computations can be carried out quite
explicitly. Eventually, the formal proof will consist in reducing the
general case to the special one.

The case discussed first amounts to the following two assumptions:
\begin{itemize}
\item  $\cC=(\{0\}\times C)''$ and the apex of $C$ is the origin
  $0\in\R^3$;
\item $\gamma\in\cL_\Gamma$ has the special form
$\gamma=i\omega^{-\frac{3}{2}}\hat{\rho}$,
  where $\rho\in\cD_\R(\R^3)$ satisfies $\rho=-\Delta\Phi$ with a rotation
  invariant function $\Phi\in C^{\infty}_\R(\R^3)$ obeying, for some
  $0<r_1<r_2<\infty$ ,
  $$ \Phi(\vx) = \left\{ \begin{array}{ll}
                      0               & \mbox{\rm if $|\vx|<r_1$},   \\
          \frac{q_\gamma}{4\pi |\vx|} & \mbox{\rm if $|\vx|>r_2$}.
  \end{array}  \right.  $$
\end{itemize}

To proceed, we note that the cone $C\subset\R^3$ determines, by projection
onto the unit sphere $S^2$,  a subset of $S^2$ which we denote by $C$,
too. Now we choose a function $\chi^C\in C^{\infty}_\R(S^2)$ with the
properties
$$ (i)\;\; \chi^C |_{S^2\setminus C} =1 \quad\quad \mbox{and} \quad\quad
   (ii)\;\; \langle Y_{00}, \chi^C \rangle =0 $$
and denote by
$\Phi^C\in C^{\infty}_\R(\R^3)$ the
product\footnote{We use the notation $\Psi\cdot\eta$ for the pointwise
             product of a rotation invariant function $\Psi$ and the
function
             $\vu\longmapsto\eta(\frac{\vu}{|\vu|})$, where $\eta\in
C^{\infty}(S^2)$. For
             definiteness, we let $(\Psi\cdot\eta) (0)\df 0$.}
$$ \Phi^C(\vx) \df (\Phi \cdot \chi^C) (\vx)
             \df \Phi(\vx) \, \chi^C(\frk{\vx}{|\vx|}).   $$
This function will now be used to construct a unitary intertwiner from
$\pi_T$ to $\pi_T\circ\gamma$ on the C$^*$-algebra $\cW(\cL(\cC'))$.

For this purpose, we calculate (using spherical coordinates)
$$ -\Delta\Phi^C = \rho \cdot \chi^C + \frac{\Phi}{r^2} \cdot
\vec{L}^2\chi^C.
$$
This function is square-integrable, hence its Fourier transform
$u^C \df - \widehat{\Delta\Phi^C}$ lies in $\cK$, and
$$ v^C_n \df  i\omega^{-\frac{3}{2}} P_{\epsilon_n}u^C, \quad n\in\N   $$
is a well-defined sequence in $D_0$ which approximates the linear form
$l_\gamma$ on $\cC'$ in the following sense:

\begin{lem}
  For any $f\in\cL(\cC')$, one has
  $l_\gamma(f)= -\lim_{n\to\infty} {\rm Im} \langle v^C_n , f \rangle$.
\end{lem}

Proof: Write
$f=\omega^{-\frac{1}{2}}\hat{h}+i\omega^{+\frac{1}{2}}\hat{g}$
with $h,g\in\cD_\R(\R^3\setminus\overline{C})$ and consider
$$ -{\rm Im} \langle v^C_n, f \rangle
   = -{\rm Im} \langle i \omega^{-\frac{3}{2}} P_{\epsilon_n} u^C,
                \omega^{-\frac{1}{2}}\hat{h}+i\omega^{+\frac{1}{2}}\hat{g}
\rangle
   = \int_{|\vk|>\epsilon_n} \!\!\!\!\!\!\!\!
           d^3k\:\omega^{-2}\overline{u^C(\vk)}\:\hat{h}(\vk).        $$
Since $u^C\in L^\infty_{\mbox{\rm \scriptsize loc}}(\R^3)$ (cf.\
Lemma~4.3),
whence $\widehat{\Phi^C}=\omega^{-2} u^C\in L^1_{\mbox{\rm \scriptsize
loc}}(\R^3)$,
it follows that this sequence converges for ${n\to\infty}$ to
$$ \int_{\R^3} d^3k \:\overline{\widehat{\Phi^C}(\vk)}\: \hat{h}(\vk)
=\Phi^C(h)
  =\Phi(h)= \int_{\R^3} d^3k \:\overline{\widehat{\Phi}(\vk)}
\:\hat{h}(\vk)
.$$
In the previous line, we have viewed $\Phi^C$ and $\Phi$ as distributions
and made use of the fact that they coincide on $\mbox{\rm supp}\!h$. The
proof is now completed by a straightforward computation showing that the
last expression equals $l_\gamma(f)$.
\Bix

Whereas property $(i)$ of $\chi^C$ was essential for the previous
Lemma, the following one will show how property $(ii)$ determines
the behaviour of $u^C$ in a neighbourhood of $\vk=0$.

\begin{lem} There exists a smooth function $\eta\in C^\infty(S^2)$ with
  $\langle Y_{00}, \eta \rangle =0$ and an analytic function
  $R:\R^3\longrightarrow\C$ with $R(0)=0$ such that
  $$ u^C(\vk) = \eta(\frk{\vk}{|\vk|}) + R(\vk)\quad\quad\mbox{for
$\vk\not=0$.} $$
\end{lem}

Proof: Let $\cS_{00}$ denote the set of all rotation invariant test
functions. Since $\langle Y_{00}, \vec{L}^2\chi^C \rangle =0$, there
exists a unique distribution $F_1$ on $\R^3$ which is homogeneous of
degree $-3$ and which coincides on $\R^3\!\setminus\!\{0\}$ with
$\frac{q_\gamma}{4\pi}\frac{1}{r^3}\!\cdot\!\vec{L}^2\chi^C$. By
Thms.~7.1.16 and 18 of \cite{Hoe} it follows that its Fourier
transform $\hat{F_1}$ is homogeneous of degree 0 and restricts on
$\R^3\setminus \{0\}$ to a smooth function, i.e.,
$\hat{F_1}(\vk) = \eta(\frk{\vk}{|\vk|})$ for $\vk\not=0$ with some
$\eta\in C^\infty(S^2)$. Moreover, since $F_1|_{\cS_{00}}=0$ and
$\cS_{00}$
is stable under Fourier transformations, it follows that $\langle
Y_{00}, \eta \rangle =0$. Now consider the distribution $F_2\df
-\Delta\Phi^C
- F_1$. For $r\not=0$, it is given by
$F_2= \rho\cdot\chi^C + (\Phi-\frac{q_\gamma}{4\pi
r})\frac{1}{r^2}\cdot\vec{L}^2\chi^C$
and thus has compact support. Hence, its Fourier transform is an
analytic function $R$: $\hat{F_2}(\vk) = R(\vk)$, $\vk\in\R^3$. As
$\chi^C$ was assumed to fulfill $\langle Y_{00}, \chi^C \rangle =0$, it
follows in particular that $R(0)= \int d^3x\,(\rho\cdot\chi^C)(\vx)=0$. To
sum up, we have $-\Delta\Phi^C = F_1 +F_2$ (in the sense of distributions)
and thus $-\widehat{\Delta\Phi^C}=\hat{F_1}+ \hat{F_2}$. Since all three
terms of this last equation are smooth on $\R^3\setminus \{0\}$, this
implies the identity
$u^C(\vk)= -\widehat{\Delta\Phi^C}(\vk)= \eta({\vk}/{|\vk|})+ R(\vk)$
for all $\vk\not=0$.
\Bix

The knowledge of $u^C$ at $\vk=0$ now permits us to establish the
connection with the KPR-like infravacuum representations described in
Section~3.

\begin{lem}
  Let $T$ be a KPR-like symplectic operator. Then:
\begin{enumerate}
  \item The limit $v_T^C\df \lim_{n\to\infty}Tv_n^C$ exists in $\cK$.
  \item The unitary $W(v_T^C)$ satisfies
$$ {\rm Ad}W(v_T^C)\circ\pi_T = \pi_T\circ\gamma \quad\mbox{on
$\dA(\cC')$}. $$
\end{enumerate}
\end{lem}

Proof: Since $\Delta\Phi^C$ is real-valued, one has $\Gamma u^C=u^C$,
hence
$\Gamma v_n^C=-v_n^C$. Thus,
$Tv_n^C = T_1 v_n^C
= i T_1\omega^{-\frac{3}{2}}P_{\epsilon_n}u^C
= i T_1\omega^{-\frac{3}{2}}P_{\epsilon_n}(u^C_1+u^C_2) $ with
$u^C_1,u^C_2\in\cK$ defined by $u^C_1 \df (1-P_0)(1\cdot\eta)$ and
$u^C_2\df u^C-u^C_1 = P_0(1\cdot\eta)+R$, where $\eta$ and $R$ are as
in the previous Lemma. In particular, $R(0)=0$ yields
$u^C_2\in D_{\omega^{-3/2}}$ which implies
$T_1\omega^{-\frac{3}{2}}P_{\epsilon_n}u^C_2 = T_1
P_{\epsilon_n}\omega^{-\frac{3}{2}}u^C_2
\stackrel{n\to\infty}{\longrightarrow}T_1\omega^{-\frac{3}{2}}u^C_2 $
by the boundedness of $T_1$. On the other hand, it follows from
$\langle Y_{00}, \eta \rangle =0$ by Lemma~3.5 that the sequence
$(T_1\omega^{-\frac{3}{2}}P_{\epsilon_n}u^C_1)_{n\in\N}$ is convergent,
which
completes the proof of Part~1. Part~2 is a
straightforward computation: Let $f\in\cL(\cC')$; then, by Lemma~4.2,
${\rm Im}\langle v_T^C, Tf\rangle = \lim_{n\to\infty}{\rm Im}\langle T
v_n^C,
Tf\rangle = \lim_{n\to\infty}{\rm Im}\langle v_n^C, f\rangle =
-l_\gamma(f) $,
which implies
\begin{align*}
  {\rm Ad}W(v_T^C)(\pi_T(W(f)))
   & =  W(v_T^C) W(Tf) W(v_T^C)^*=
           e^{-i\mbox{\footnotesize ${\rm Im}$}\langle v_T^C,
Tf\rangle}W(Tf)
\\
   & =  e^{i l_\gamma(f)}W(Tf ) = \pi_T\circ\gamma (W(f)).
\end{align*}
This establishes the stated equivalence on $\cW(\cL(\cC'))$ and, hence,
by local normality (of both $\pi_T$ and $\gamma$) also on $\dA(\cC')$.
\Bix

With these preparations, we are ready for the \\
Proof of Prop.~4.1:
By standard arguments using transportability of the charges,
$\gamma_x\cong\gamma$, and translation covariance of the representation
$\pi_T$, it can always be assumed that the apex of $\cC$ is
$0\in\R^{1+3}$. To remove the assumption on the special form of $\gamma$
as
well, we note that, for any $\gamma\in\cL_\Gamma$, there exists some
equivalent
$\gamma_0\in\cL_\Gamma$ with the special form considered. Such a
$\gamma_0$
automatically satisfies $\gamma_0-\gamma\in D_T$ and thus provides a
unitary $W(T(\gamma_0-\gamma))$ performing the equivalence
$\pi_T\circ\gamma\cong\pi_T\circ\gamma_0$ on all of $\dA$. Taking into
account
all the above, this proves Prop.~4.1
\Bix

\section{Conclusions}
The present work has shown in a concrete example that choosing a
background different from the vacuum can improve the localisability
properties of superselection sectors in theories with massless
particles. Typically, such backgrounds correspond to clouds of infrared
radiation which exist in great variety in any such theory, but it
appears that {\em suitable} background states have to be chosen carefully
in order to match with the sectors under consideration. To illustrate
this point, we recall that this led us, in particular,
to  choose complex conjugation in {\em position} space as
the involution $\Gamma$. If we had, on the other hand, chosen complex
conjugation $\hat{\Gamma}$ in {\em momentum} space instead of $\Gamma$,
our
sectors would only have been localisable in  {\em upright spacelike
double cones}, i.e., in regions of the form $\cC\cup(a-\cC)$, where $\cC$
is an upright spacelike cone and $a\in\R^{1+3}$,  but not in spacelike
cones 
any more. (The former statement can be verified with the method of
Section~4, 
the latter can be reduced to an application of Lemma~2.2.)

Finally, we remark  that we were unable, with our methods, to
establish the {\em full} BF localisation criterion for the sectors
$[\gamma]$,
i.e.,  $ \pi_T\circ\gamma|_{\dA(\cC')} \cong \pi_T |_{\dA(\cC')} $ also
for spacelike cones $\cC$ which do not contain an upright one. Apart
from the obvious fact that the KPR-like states break Lorentz
covariance explicitly, there seem to be other indications that such
a result  might indeed not be true. However, we do not pursue this point
further, since the localisation properties obtained here should be
sufficient for carrying through a DHR-like analysis (along the lines
of \cite{BF82}) in front of the infravacuum background. If this is
indeed possible, it has to be studied in  a subsequent step under
which conditions (to be imposed on the infravacuum) the superselection
structure thus obtained will be independent from the particular
background.

\vspace{2em} \noindent
{\bf Acknowledgements:} I would like to thank Prof.\ D.\ Buchholz for
numerous helpful discussions and stimulating interest in this
work. Partial financial support from the Deutsche Forschungsgemeinschaft
(``Graduiertenkolleg Theoretische Elementarteilchenphysik'' at Hamburg
University) is also gratefully acknowledged.

\end{document}